%% file: main.tex
\documentclass[journal]{IEEEtran}
\ifCLASSINFOpdf
\else
\fi
\usepackage{amsmath,amsthm}
\usepackage{amssymb,amsfonts}
\usepackage{algorithmic}
\usepackage{graphicx}
\usepackage{array}
\usepackage{cancel}
\usepackage[linesnumbered,algoruled,boxed,lined]{algorithm2e}
\usepackage{textcomp}
\usepackage{comment}

\usepackage{times}  %Required
\usepackage{helvet}  %Required
\usepackage{todonotes}
\usepackage{bm}
\usepackage{orcidlink}
\usepackage{tikz}
\usetikzlibrary{arrows,automata}
\usepackage{float}
\usepackage{subcaption}
\usepackage{mathtools}
\usepackage{acronym}
\usepackage{courier}  %Required
\usepackage{url}
\usepackage{colonequals}
\usepackage{graphicx}  %Required
\input{defs.tex}
\usepackage{paralist}

\usepackage{flushend}

\begin{document}
\title{Dynamic Hypergames for Synthesis of Deceptive Strategies with Temporal Logic Objectives}

\author{Lening Li \orcidlink{0000-0002-8321-1254},~\IEEEmembership{Student Member,~IEEE,}
Haoxiang Ma \orcidlink{0000-0002-3823-385X},~\IEEEmembership{Student Member,~IEEE,}
Abhishek N. Kulkarni \orcidlink{0000-0002-1083-8507},~\IEEEmembership{Student Member,~IEEE,}
and
Jie Fu \orcidlink{0000-0002-4470-2827},~\IEEEmembership{Member,~IEEE}% <-this % stops a space
\thanks{}% <-this % stops a space
\thanks{Lening Li, Haoxiang Ma, and Abhishek N. Kulkarni are with Robotics Engineering Program, and Jie Fu is with Faculty of Electrical and Computer Engineering, with affiliation to Robotics Engineering Program, Worcester Polytechnic Institute, Worcester, MA. 01609. e-mail: lli4, hma2,  ankulkarni, jfu2@wpi.edu.}%
\thanks{This material is based upon work supported by the Defense Advanced Research Projects Agency (DARPA) under Agreement No. HR00111990015. Approved for public release; distribution is unlimited.}}

\maketitle

\begin{abstract}
In this paper, we study the use of deception for strategic planning in adversarial environments. We model the interaction between the agent (player 1) and
  the adversary (player 2)  as a two-player concurrent game
  in which the adversary has incomplete information about the agent's
  task specification in temporal logic. During the online interaction, the adversary can
  infer the agent's intention from  observations and adapt its
  strategy so as to prevent the agent from satisfying the task.  To plan against such an adaptive opponent, the agent must leverage its knowledge about  the adversary's incomplete information to influence the behavior of the opponent, and thereby being deceptive. To synthesize a deceptive strategy, we introduce a class of hypergame models that capture the  interaction between the agent and its adversary given asymmetric, incomplete information. A hypergame is a hierarchy of games, perceived
  differently by the agent and its adversary. We develop the solution
  concept of this class of hypergames and show that the subjectively rationalizable strategy for the agent is deceptive and maximizes the probability of satisfying the task in temporal logic.  This deceptive strategy is obtained by 
  modeling the opponent evolving perception of the interaction and integrating the opponent model into proactive planning. Following the deceptive strategy, the agent
  chooses actions to influence the game history as well as to
  manipulate the adversary's perception so that it takes actions that
  benefit the goal of the agent.  We demonstrate the correctness of
  our deceptive planning algorithm using robot motion planning
  examples with temporal logic objectives and design a detection mechanism to notify the agent of potential errors in modeling of the adversary's behavior.
\end{abstract}

\begin{IEEEkeywords}
Hypergame, Markov Decision Processes, Linear Temporal Logic, Deception.
\end{IEEEkeywords}

\def\abstractname{Note to Practitioners}
\begin{abstract}
 Many security and defense applications employ deception mechanisms for strategic advantages. This work presented a game-theoretic framework for planning deceptive strategies in stochastic environments and shows that the opponent modeling plays a key role in the design of effective deception mechanisms. For applications to cyber-physical security, the practitioners can employ temporal logic for specifying security properties in the system and analyze defense with deception using the proposed methods.
 \end{abstract}

% \fi
\IEEEpeerreviewmaketitle

\section{Introduction}
\label{sec:introduction}
\input{sections/introduction.tex}

\section{Preliminaries and Problem Formulation}
\label{sec:preliminaries}
\input{sections/preliminaries.tex}

\section{Formulating a class of omega-regular hypergame}
\label{sec:hypergame}
\input{sections/hypergame.tex}

\section{Case study}
\label{sec:case_study}
\input{sections/case_study}

\section{Conclusion}
\label{sec:conclusion}
\input{sections/conclusion.tex}
\ifCLASSOPTIONcaptionsoff
  \newpage
\fi



%\bibliographystyle{IEEEtran}


% \end{thebibliography}
\bibliographystyle{IEEEtran}
\bibliography{refs.bib}


\begin{IEEEbiography}[{\includegraphics[width=1in,height=1.25in,clip,keepaspectratio]{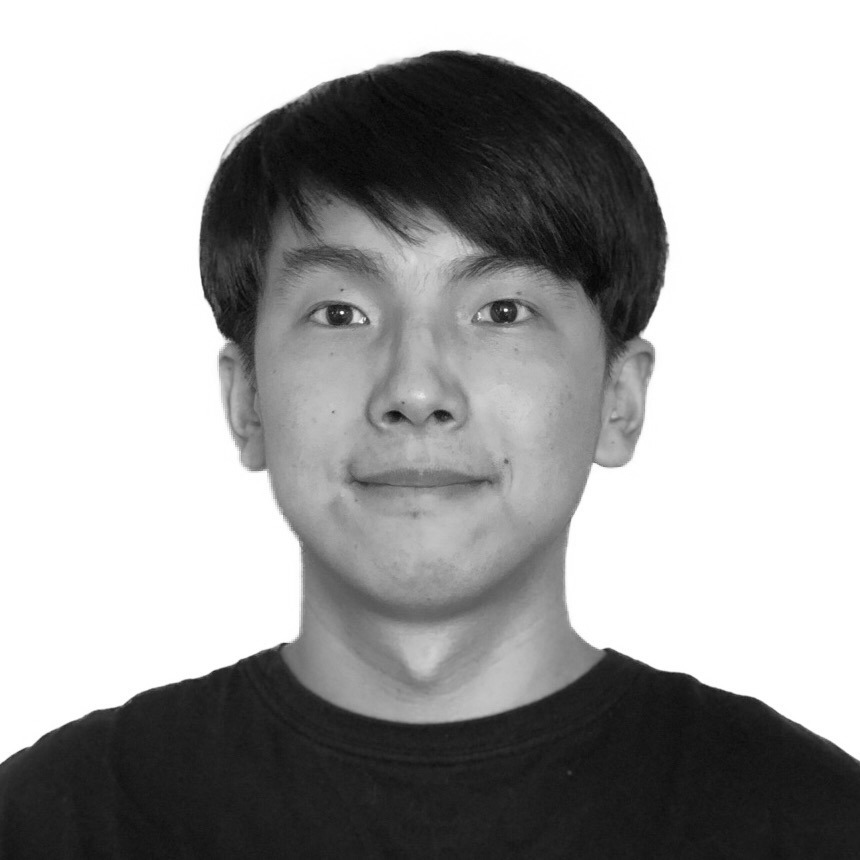}}]{Lening Li} (Student Member,~IEEE)
received a B.S. degree in Information Security (Computer Science) from Harbin Institute of Technology, China in 2014, and an M.S. degree in Computer Science from Worcester Polytechnic Institute, Worcester, MA, USA, in 2016. He is currently pursuing a Ph.D. degree in Robotics Engineering at Worcester Polytechnic Institute, Worcester, MA, USA. 

His research interests include reinforcement learning, stochastic control, game theory, and formal methods.
\end{IEEEbiography}

\begin{IEEEbiography}[{\includegraphics[width=1in,height=1.25in,clip,keepaspectratio]{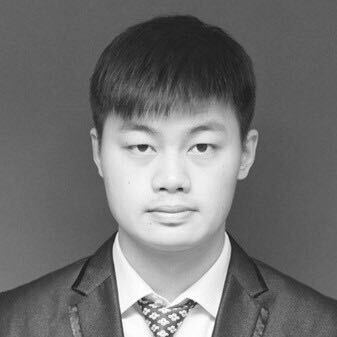}}]{Haoxiang Ma} (Student Member,~IEEE)
received the B.S. degree in Computer Science from the Nankai University, China in 2017, and the M.S. degree in Computer Science from Worcester Polytechnic Institute, Worcester, MA, USA, in 2019. He is currently pursuing the Ph.D. degree in Robotics Engineering at Worcester Polytechnic Institute, Worcester, MA, USA. 

His interests include: Reinforcement Learning, Optimal control, Game theory, with applications to robotic systems.
\end{IEEEbiography}

\begin{IEEEbiography}[{\includegraphics[width=1in,height=1.25in,clip,keepaspectratio]{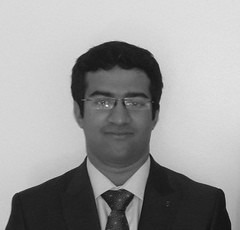}}]{Abhishek N. Kulkarni} (Student Member,~IEEE) received the B.Tech. degree in Electronics and Telecommunications Engineering from the Vishwakarma Institute of Technology, India in 2016. He is currently pursuing his Ph.D. degree in Robotics Engineering at Worcester Polytechnic Institute, Worcester, MA, USA.

His interests include: game and hypergame theory, formal methods with applications to robotic systems and cybersecurity. 

\end{IEEEbiography}

\begin{IEEEbiography}[{\includegraphics[width=1in,height=1.25in,clip,keepaspectratio]{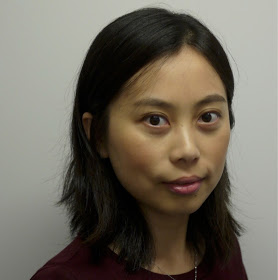}}]{Jie Fu} (Member,~IEEE) received the B.S. and M.S. degrees
from Beijing Institute of Technology, Beijing, China,
in 2007 and 2009, respectively, and the Ph.D. degree
in mechanical engineering from the University of
Delaware in 2013. She is currently an Assistant
Professor at the Department of Electrical and Computer Engineering, Robotics Engineering Program, Worcester Polytechnic Institute.

Her research interests include   probabilistic decision-making, stochastic control,  game theory, and formal methods.
\end{IEEEbiography}

%\newpage



%\vfill


\end{document}

%% file: defs.tex
\newif\ifuseboldmathops
\newif\ifuseittextabbrevs
\useboldmathopstrue   % comment out to use mathbb
\ifuseittextabbrevs
  
  \newcommand{\eg}{{\it e.g.}}
  \newcommand{\ie}{{\it i.e.}}

\else
  
  \newcommand{\eg}{e.g.}
  \newcommand{\ie}{i.e.}

\fi

\ifuseboldmathops
  \newcommand{\reals}{\mathbf{R}}

\else
  \newcommand{\reals}{\mathbb{R}}

\fi

\ifuseboldmathops

\else

\fi

\ifuseboldmathops

\else

\fi

\ifuseboldmathops

\else

\fi

\newcommand{\Always}{\Box \, }
\newcommand{\Eventually}{\Diamond \, }

\newcommand{\until}{\mbox{$\, {\sf U}\,$}}

\newcommand{\sink}{\mathsf{sink}}

\newcommand{\BoolTrue}{\mbox{\sf true}}

\newcommand{\abs}[1]{\lvert#1\rvert}

\newcommand{\dist}[1]{\mathcal{D}(#1)}

\newcommand{\calF}{\mathcal{F}}
\newcommand{\calAP}{\mathcal{AP}}

\newcommand{\indicator}{\mathbf{1}}

\renewcommand{\Pr}{\mathbf{Pr}}

\newcommand{\calA}{\mathcal{A}}
\newcommand{\game}{\mathcal{G}}
\newcommand{\hgame}{\mathcal{HG}}

\newcommand{\plays}{\mathsf{Plays}}
\newcommand{\prefplays}{\mathsf{PrefPlays}}

\theoremstyle{definition}
\newtheorem{definition}{Definition}

\newtheorem{problem}{Problem}
\newtheorem{assumption}{Assumption}

\newtheorem{remark}{Remark}
\newtheorem{theorem}{Theorem}[section]
\newtheorem{lemma}[theorem]{Lemma}

\acrodef{mdp}[MDP]{Markov Decision Process}
\acrodef{pomdp}[POMDP]{Partially Observable Markov Decision Process}
\acrodef{ltl}[LTL]{Linear Temporal Logic}
\acrodef{dfa}[DFA]{Deterministic Finite Automaton}
\acrodef{scltl}[scLTL]{syntactically co-safe LTL}
\acrodef{ssp}[SSP]{Stochastic Shortest Path}
\acrodef{mcts}[MCTS]{Monte Carlo tree search}
\acrodef{mc}[MC]{Markov chain}
\acrodef{sr}[SR]{subjectively rationalizable}

\acrodef{bsr}[BSR]{behaviorally subjectively rationalizable}
\acrodef{cusum}[CUSUM]{Cumulative SUM}

\newcommand{\NE}{\dagger}

%% file: sections/introduction.tex
\IEEEPARstart{P}{lanning} in adversarial environments is commonly encountered in security
and defense applications. When the objective of a planning agent is
partially unknown to its adversary, deception becomes inseparable from
the strategic planning: With a proper deception mechanism, the
adversary can be misled to take actions beneficial to the
agent. Scenarios of interest include
economics~\cite{gneezy2005deception}, military
operations~\cite{handel2005masters}, securities in
cyber-network~\cite{stech2016integrating,horak2017manipulating,Huang2018,Ahmadi2018ThePO},
robotics, and other cyber-physical
systems~\cite{wagner2011acting,egorov2016target,Masters2017,gharesifard2012evolution,ornik2018deception}.
This paper investigates deceptive planning by developing solution
concepts for a class of concurrent stochastic games with asymmetric
information and Boolean payoffs in temporal
logic~\cite{manna2012temporal}. In this game, the agent (player 1/P1) is to
achieve a task specified as a temporal logic
formula~\cite{kupferman2001model}, which expresses desired behaviors
including safety, reachability, obligation, and liveness. The
asymmetry of information is introduced such that the adversary (player
2/P2) does not know player 1's task formula
but  is aware of its incomplete information about player 1. The question is, how could player 1 leverage player 2's incomplete information for optimizing task performance?

To address this question, we take a game-theoretic approach. In literature, deceptive planning has employed the solution concepts of games with
asymmetrical information including dynamic Bayesian
games~\cite{Huang2018,zhang2019game} and
hypergames~\cite{house2010hypergame,ferguson2019game}. The
authors~\cite{Huang2018} adopted dynamic Bayesian games to defensive planning in cyber
deception, where the type of the opponent, \ie, a legitimate user or an
attacker, was the private information. They captured the incomplete
information as a probability distribution over a set of the opponents'
types and employed Bayesian Nash Equilibrium to design a defensive
strategy. Hypergame \cite{wang1989solution,bennett1980hypergames}
was characterized by players' misperception of the other players'
payoffs or other components of the game (state/action space). In
repeated normal-form hypergames, the authors
\cite{gharesifard2012evolution} studied how a player's belief of other
players' preferences evolves  by observing other players' decisions and
analyzing the inconsistency in the equilibrium. They developed
stealthy deception~\cite{gharesifard2014stealthy} to restrict the
deceiver's actions so as not to contradict the
belief of the deceivee. Deception has been also investigated for enhancing
security of cyber-physical systems (CPSs). In
\cite{sayin2019deception}, the authors
proposed a deception-as-defense framework to ensure optimal system
performance given adversarial inputs in communication and control of
continuous systems. The use of deception is to allow the agent with
private information to craft the information perceived by adversary in order
to control adversary's perception of the state of the system. For the
supervisory control of CPSs, the authors
\cite{goes2017stealthy} developed an algorithm to
synthesize an attack strategy with which the attacker modifies sensor
readings and thus misleads the supervisor to achieve undesirable
states in the discrete event systems.  It is noted that these deception methods in CPSs
are concerned with deceptive information exchange, instead of the
payoffs about the game. Besides game-theoretic approaches, deceptive  planning was developed in \cite{Masters2017,ornik2018deception} when the deceiver hides its objective from the observer for achieving the goal. However, there is no dynamic interaction between the deceiver and the observer.

A major characteristic in our game model is that players' payoffs are temporal logic formulas, instead of utility functions. In reactive synthesis, games with temporal logic payoffs, also known as omega-regular games, are studied for synthesizing  provably correct systems in dynamic environments. Existing solution concepts for omega-regular games studied games with complete information and perfect/partial observations \cite{chatterjee2012survey,bloemGraphGamesReactive2018,dealfaroConcurrentReachabilityGames2007,pitermanSynthesisReactiveDesigns2006}. 
However, to synthesize deceptive strategy given temporal logic tasks, it is necessary to develop new  theory and solution concepts of omega-regular games with asymmetric, incomplete information.

% Besides game-theoretic approaches that involve two
% agents interacting, deceptive path-planning was developed in \cite{Masters2017}
% and considers a deceptive agent and an observer. The goal was to
% minimize the probability of an observer identifying the agent's final
% destination before the final destination had been reached. Similar
% idea has been seen in \cite{ornik2018deception}, where one agent
% sought to achieve an objective while confusing the opponent about the
% true intention by increasing the entropy in the planned policy. The
% authors proposed multi-objective optimal planning to solve the
% deceptive policies in stochastic environments.

To this end, we propose both a modeling framework and synthesis
methods for deceptive planning in a subclass of omega-regular games
with hierarchical information pattern. In this subclass of
omega-regular games, players' objectives are given by \ac{scltl} formula and safe \ac{ltl} formulas \cite{klein2006experiments}. We extend
hypergame definition to define a class of hierarchical omega-regular
hypergames.  
% A hierarchical hypergame is a meta-game (or game of
% games), that models the perception of the game of individual players,
% as well as the perception of other player's perception of the game
% given private information. We assume that  the adversary has
% \begin{inparaenum}[1)]
%   \item  a correct
%   model about the interaction dynamics with the agent but incomplete
%   information about the agent's intentions,
%   \item the ability of making
%   rationalizable decisions given his/her perception, and
%   \item
%   the ability of inferring (part of) the agent's intention from
%   observations at runtime.
% \end{inparaenum} The goal of the agent is to maximize the probability of satisfying the logical specification. This is achieved through joint
% planning in the game state, task state, and information state of the
% adversary.
% The deceptive strategy is obtained from the solution concept of
% hierarchical omega-regular hypergames. In particular, 
We develop a
solution concept called subjectively rationalizable strategies in
hypergame. Based on this solution, two key modules, namely, \emph{opponent
  modeling} and \emph{proactive planning}, are identified for effective
deception. Given the common knowledge that a temporal logic formula
describes a sequence of temporally extended subgoals, we assume that
the adversary can infer the current agent's subgoal based on
observations.  In opponent modeling, the agent maintains a model of
the adversary's subjectively rationalizable behavior strategy and
subgoal inference. Using the opponent model, the agent predicts the adversary's
subjectively rationalizable strategy and plans proactively to satisfy its temporal logic constraints. The term ``proactive" means that the agent predicts how its action will influence the perception of the
adversary before taking the action. The effectiveness of the proactive strategy hinges on the
matching between the agent's opponent model and the true opponent. We
 design an online detection algorithm to identify potential errors in the
opponent model. We show the effectiveness of
the proposed solution concept for deceptive planning using robot
motion planning examples with temporal logic objectives.

To summarize, the key contributions of the paper are:
\begin{IEEEitemize}
  \item A modeling framework of hierarchical omega-regular hypergames
        for adversarial interactions with asymmetric information about the
        payoffs in temporal logic.
  \item A solution concept for this class of hypergames.
  \item A proactive deceptive planning method that integrates
        opponent modeling, intent inference, and probabilistic planning with temporal
        logic constraints.
  \item A detection mechanism to identify at runtime any modeling mismatch that may make the deceptive strategy ineffective.
\end{IEEEitemize}

%% Structural description of the paper
The remainder of this paper is organized as follows. Section~\ref{sec:preliminaries} provides some necessary background on game theory and linear temporal logic. Section~\ref{sec:hypergame} presents the main theory and algorithms for deceptive planning. Section~\ref{sec:case_study} presents a case study to demonstrate the effectiveness of the deception. Finally, Section~\ref{sec:conclusion} concludes and discusses future work.

%% file: sections/preliminaries.tex
\textbf{Notation:} We use the notation $\Sigma$ for a finite set of symbols, also known as the alphabet. A sequence of symbols $w=\sigma_0 \sigma_1 \ldots \sigma_n$ with $\sigma_i\in \Sigma$ for any $0 \leq i \leq n$, is called a \emph{finite word}, and $\Sigma^\ast$ is the set of all finite words that can be generated with alphabet $\Sigma$. The \emph{empty word}, denoted by $\varepsilon$, is the empty sequence. We denote the set of all $\omega$-regular words as $\Sigma^\omega$ obtained by concatenating the elements in $\Sigma$ infinitely many times. The length of a word is denoted by $\abs{w}$ (note that $\abs{\varepsilon} = 0$). We define the set of nonempty finite words $\Sigma^{+} = \Sigma^{\ast} \setminus \{\varepsilon\}$. 
% Given a sequence $w$, the function $\Occ(w) = \{\sigma\mid \exists 0\le i\le \abs{w}, w_i=\sigma \}$ --- outputs a set of symbols occurring in $w$.
Given a finite and discrete set $X$, let $\dist X$ be the set of all probability distributions over $X$.

\subsection{Omega-regular games}
We consider an adversarial  encounter between two players; a controllable player P1 (pronoun ``he") and an uncontrollable player P2 (pronoun ``she"). Both players choose their moves simultaneously. The dynamics of their interaction can be captured as a  transition system with simultaneous moves.

{\definition[Two-player Transition System with Simultaneous Moves]A \emph{two-player transition system with simultaneous moves} is a tuple
$
  TS = \langle S, A, P, s_0, \calAP, L \rangle
$ consisting of the following components:
\begin{IEEEitemize}
  \item $S$ is the set of states.
  \item $A = A_1 \times A_2$ is a finite set of actions, where $A_1$ is the set of actions that P1 can perform, and $A_2$ is the set of actions that P2 can perform.
  \item $P \colon S \times A \to \dist S$ is a probabilistic transition function. At every state $s \in S$, P1 chooses an action $a_1 \in A_1$, and P2 chooses an action $a_2 \in A_2$ simultaneously. Then, a successor state $s'$ is determined by the probability distribution $P(\cdot \mid s,a)$, where $a = (a_1, a_2) \in A$.
  \item $s_0$ in an initial state.
  \item $\calAP$ is a set of atomic propositions.
  \item $L \colon S \to 2^{\calAP}$ is a labeling function. For every state $s \in S$, the label $L(s)$ of the state $s$ represents a set of atomic propositions that are evaluated true at the state $s$.
\end{IEEEitemize}
}

A
\emph{path} $\rho=s_0s_1 \ldots $ is a state sequence such that for any $i\ge 0$, there
exists $a\in A$, $P(s_{i+1} \mid s_i,a) >0 $.
A path $\rho=s_0s_1 \ldots $ can be mapped to a word in $2^\calAP$,
$w=L(s_0)L(s_1) \ldots $, which is evaluated against
logical formulas.  In this paper, unless otherwise noted, we will refer to a two-player transition system with simultaneous moves simply as a transition system.

We use linear temporal logic (LTL) formulas to represent the players' objectives/payoffs in the game. Temporal Logic provides a succinct way to represent temporal goals and constraints. An \ac{ltl} formula is defined inductively as follows:
\[
  \varphi \colonequals \top \mid \bot \mid p \mid \varphi \mid \neg\varphi \mid \varphi_1 \land \varphi_2 \mid \bigcirc \varphi \mid \varphi_1 {\until} \varphi_2,
\]
where $\top$ and $\bot$ are universally true and false, respectively, $p \in \calAP$ is an atomic proposition, and $\bigcirc$ is a temporal operator called the ``next'' operator. $\bigcirc \varphi$ is evaluated to be true  if the formula $\varphi$ becomes true at the next time step. $\until$ is a temporal operator called the ``until'' operator. The formula $\varphi_1\until \varphi_2$ is true given that $\varphi_2$ will be true in some future time steps, and before that $\varphi_1$ holds true for every time step.

The operators $\Eventually$ (read as eventually) and $\Always$ (read as always) are defined using the operator $\until$ as follows: $\Eventually \varphi = \top \until \varphi$ and $\Always \varphi = \neg \Eventually \neg \varphi$. The formula $\Eventually \varphi$ is true if $\varphi$ becomes true in some future time. The formula $\Always \varphi$ means that $\varphi $ will hold for every time step. Given an \ac{ltl} formula $\varphi$ and a word $w\in \Sigma^\omega$, if the word $w$ satisfies the formula $\varphi$, then we denote $w\models \varphi$. We denote a set of \ac{ltl} formulas as $\Phi$. For details about the syntax and semantics of \ac{ltl}, the readers are referred to \cite{Pnueli1989}.

In this paper, we restrict the specifications of the players to the class of \ac{scltl} \cite{kupferman2001model}. An \ac{scltl} formula contains only $\Eventually$ and $\until$ temporal operators when written in a positive normal form (\ie, the negation operator $\neg$ appears only in front of atomic propositions). The unique property of \ac{scltl} formulas is that a word satisfying an \ac{scltl} formula $\varphi$ only needs to have a \emph{good prefix}. That is, given a good prefix $w \in \Sigma^\ast$, the word $ww' \models \varphi$ for any $w'\in \Sigma^\omega$. The set of good prefixes can be compactly represented as the language accepted by a \ac{dfa} defined as follows:
\begin{definition}[Deterministic Finite Automaton]
  Given P1's objective expressed as an \ac{scltl} formula $\varphi$, the set of good prefixes of words corresponding to $\varphi$ is accepted by a \ac{dfa} ${\cal A} = \langle Q, \Sigma, \delta, \iota, F \rangle
  $ with  the following components:
  \begin{IEEEitemize}
    \item $Q$ is a finite set of states.

    \item $\Sigma = 2^{\calAP}$ is a finite set of symbols.

    \item $\delta \colon Q \times \Sigma \to Q$ is a deterministic transition function.

    \item $\iota \in Q$ is the unique initial state.

    \item $F \subseteq Q$ is a set of final states.
  \end{IEEEitemize}
\end{definition}
For an input word $w =\sigma_0\sigma_1\ldots \in \Sigma^\omega$, the \ac{dfa} generates a sequence of states $q_0q_1\ldots $ such that $q_0=\iota$ and $q_{i+1}= \delta(q_{i},\sigma_{i})$ for any $i\ge 0$. The word $w$ is accepted by the \ac{dfa} if and only if there exists $k \ge 0$ such that $q_k \in F$. The set of words accepted by the \ac{dfa} $\calA$ is called \emph{its language}. 
We assume that the \ac{dfa} is complete --- that is, for every state-action pair $(q,\sigma)$, $\delta(q,\sigma)$ is defined.  An incomplete DFA can be made complete by adding a sink state $q_\sink$ such that $\delta(q_\sink, \sigma)=q_\sink$ and directing all undefined transitions to the sink state $q_\sink$.

In the interaction between P1 and P2, P1 aims to maximize the probability of satisfying his objective in temporal logic, say $\varphi_1$, over the transition system; P2's objective is to prevent P1 from satisfying his objective. Hence, P2's objective can be denoted by $\varphi_2 = \neg \varphi_1$. Next, we give the definition of zero-sum game with Boolean payoffs expressed in temporal logic.

\begin{definition}[Omega-Regular Game]
  \label{def:games}
  Let $\varphi_1$ denote the objective of P1. Then, given a transition system $TS$, a zero-sum omega-regular game on a graph is the tuple
  \[
    \game(\varphi_1, \neg \varphi_1) = \langle TS, \varphi_1, \neg \varphi_1 \rangle.
  \]
\end{definition}
As the objective of P2 is the logical negation of P1's objective, we omit P2's objective $\neg \varphi_1$ from $\game(\varphi_1, \neg \varphi_1)$ and simply denote the game as $\game(\varphi_1)$. A path $\rho \in S^\omega$ of $TS$ is winning for player $i$ if and only if $L(\rho)\models \varphi_i$ --- that is, the labeling of that path satisfies player $i$'s Boolean objective in temporal logic.

A \emph{play} in the game is constructed as follows: The players start with the initial game state $s_0$, simultaneously select a pair of actions $(a_1,a_2)\in A$, move to a next state $s_1$, and iterate. The game ends when one of the players satisfies its objective. Thus, a play is a sequence of states and actions, denoted by $s_0 a_0 s_1 a_1\ldots $ such that $P(s_{i+1} \mid s_i,  a_i)>0$ for any $ i\ge 0$. The set of plays in the game is denoted by $\plays$. The set of prefixes of plays ending in a state in the game is denoted by $\prefplays$. We refer to $h \in \prefplays$ as a history in the game. Given a history $h$, we denote the projection of $h$ onto the set of states $S$ as $h\downharpoonright_{S}$.

A (mixed) strategy $\pi_i \colon \prefplays \to \dist {A_i}$, for player $i \in\{1,2\}$, is a function that assigns a probability
distribution over all actions given a history.  Let $\Pi_i$ denote the (mixed) strategy space of player $i$. A strategy profile $\langle \pi_1, \pi_2 \rangle$ is a pair of strategies, one for each
 player. A strategy profile $\langle \pi_1, \pi_2 \rangle$ induces a probability measure $\Pr^{\langle \pi_1,\pi_2 \rangle}$ over $\prefplays$.
 
Slightly abusing the notation, given a play $h \in \plays$, we say  that $h\models \varphi$ for an \ac{ltl} formula $\varphi$ if $L(h\downharpoonright_S)\models \varphi$, that is, the sequence of states labels obtained from the projection of $h$ onto $S$ satisfies the \ac{ltl} formula $\varphi$. %, where the history $h$ in the game is a realization of the stochastic process $(S_t)_{t\ge 0}$ under the probability distribution $\Pr^{\langle \pi_1,\pi_2 \rangle}$.

% The P1's strategy $\pi_1$ is \emph{almost-sure winning} given a history $h$ if for every P2's strategy $\pi_2\in \Pi_2$, $\Pr^{\langle\pi_1,\pi_2 \rangle}(h \models \varphi_1)=1$.

Given player $i$'s Boolean objective $\varphi_i$, we define the utility function for player $i$ as
$u_i\colon \prefplays(\game) \times \Pi_i\times \Pi_j\times \Phi \to \reals$, such that for $(i, j) \in \{(1, 2), (2,1)\}, u_i(h, \pi_i,\pi_j, \varphi_i)= \Pr^{\langle \pi_i,\pi_j \rangle}(hh' \models \varphi_i)$ is the probability of satisfying the specification $\varphi_i$, where $h \in \prefplays$ is the initial history given that players follow the strategy profile $\langle \pi_i,\pi_j \rangle$, and $h'$ is the stochastic process induced by strategy profile $\langle \pi_i,\pi_j \rangle$ after the initial history $h$. 
 
We present the definition of Nash equilibrium for omega-regular games with complete information as follows. 
 \begin{definition}[Nash equilibrium~\cite{osborne1994course}]
A Nash equilibrium of a omega-regular game $\game(\varphi_1)$ is a strategy profile $\langle \pi^\ast_1, \pi^\ast_2 \rangle$ with the property that for $(i, j) \in \{(1, 2), (2,1)\}$ we have 
\[
    u_i(h, \pi_i^\ast, \pi_j^\ast, \varphi_i) \ge u_i(h, \pi_i, \pi_j^\ast, \varphi_i).
 \]
\end{definition}

In zero-sum omega-regular game, the Nash equilibrium  $\langle\pi_1^\ast, \pi_2^\ast\rangle$ can be obtained as follows:
\begin{equation}
  \label{eq:optimal-strategy}
  \langle \pi^\ast_1, \pi^\ast_2\rangle=\arg\max_{\pi_1\in \Pi_1}\min_{\pi_2\in \Pi_2}\Pr^{\langle \pi_1,\pi_2 \rangle}(hh' \models\varphi_1 ).
\end{equation}

\subsection{Problem formulation: Planning under Information Asymmetry}
In Def.~\ref{def:games}, the game is a common knowledge to both players. We now consider the case when the information about the game (\eg, dynamics, payoffs) between two players is asymmetrical. Specifically, we consider the case when P2 has incomplete information about P1's  temporal logic objective.
\begin{assumption}\label{assume:knowledge}
  The asymmetrical information between players is introduced as follows:
  \begin{IEEEitemize}
    \item P1's objective is $\varphi_1$.
    \item P2 does not know $\varphi_1$ but has an initial hypothesis $x_0$ and a hypothesis space $X$ about P1's objective.
  \end{IEEEitemize}
\end{assumption}
The assumption describes scenarios commonly encountered in practice, for both cooperative and adversarial interactions. The problem we aim to solve is stated informally as follows: 

\begin{problem}
Given an adversarial encounter between P1 and P2 under information asymmetry as defined by Assumption~\ref{assume:knowledge}, how to compute a best-response strategy for P1 that maximizes the probability of satisfying $\varphi_1$ while a rational P2 responds optimally given P2's knowledge of the game?
\end{problem}

The definition of ``best-response'' differs from the one for games with symmetric and complete information in equation~\eqref{eq:optimal-strategy}. Next, we introduce the modeling framework of hypergames and present a solution concept for a class of hypergames to solve P1's strategy.

%% file: sections/hypergame.tex
Hypergame, introduced in \cite{bennett1980hypergames}, is  capable of modeling strategic interactions when players have asymmetrical information. Intuitively, a hypergame is a game of games, and each game is associated with a player's subjective view of its interaction with other players based on its own information and information about others' subjective views.

\subsection{Static hypergames on graphs}
To start with, given that P2 has incomplete information, we introduce a \emph{hypothesis space} for P2, denoted by $X$. The set $X$ can be discrete and finite. For example, the set $X$ can be a finite set of \ac{scltl} formulas that P2 believes that P1's true objective is one of these. The hypothesis space $X$ can also be continuous. For example, each $x \in X$ is a distribution over a subset of \ac{scltl} formulas $\Phi$ so that $x(\varphi)$ is the probability that P2 believes $\varphi$ to be P1's true objective. For the time being, we do not restrict the set $X$.

We can formally introduce a hypergame, which extends the hypergames from normal-form games \cite{bennett1980hypergames,Vane} to omega-regular games. 
\begin{definition}
  A static omega-regular hypergame of level-1 is defined as \[{\hgame}^1 (x)= \langle \game(\varphi_1), \game(x) \rangle,\] where $\game(\varphi_1)$ is the game constructed by P1 given his objective $\varphi_1$, and $\game(x)$ is the game constructed by P2 given her hypothesis $x \in X$. If P1 is aware of P2's game $\game(x)$, then the resulting hypergame is said to be of level-2 and defined as
  \[
    {\hgame}^2 (x)= \langle {\hgame}^1(x), \game(x) \rangle.
  \]
  In a level-2 hypergame, where $\hgame^1(x)$ is the level-1 hypergame constructed by P1 given his objective and his knowledge about the game constructed by P2, P1 computes his strategy by solving the level-1 hypergame $\hgame^1(x)$ while P2 computes her strategy by solving the game $\game(x)$. 
\end{definition}
The game constructed by a player given his/her information and higher-order information is called the player's perceptual game. In level-2 hypergame $\hgame^2(x)$, P1's perceptual game is $\hgame^1(x)$ and P2's perceptual game is $\game(x)$.

If the hypothesis $x \in X$ does not change, then we say this game is static. In this static hypergame, players choose actions simultaneously, determined by their respective strategies, and their strategies do not change during their interaction.   

Higher levels of hypergames can be defined through a recursive reasoning about higher-order information (\eg, what I know that you know that I know ...).
In this scope, we restrict to level-2 hypergames.
For simplicity, we refer to level-2 omega-regular hypergames  as hypergames in this paper whenever it is clear from the context.

We now discuss the solution concepts of hypergames. Given that different players may have different perceptions (\ie, subjective views) of the utility functions in a hypergame, we denote $u_i^j$ as the \emph{utility function of player $i$ perceived by player $j$}. 
Next, we generalize the related notions of subjective rationalizability and best-response equilibrium in hypergames from normal-form games in \cite{sasaki2016hierarchical} to omega-regular hypergames.
\begin{definition}[Subjective Rationalizability]
  Given a level-2 hypergame $\hgame^2(x) = \langle \hgame^1(x), \game(x)  \rangle$, strategy $\pi_i^{\ast,2}$ is \ac{sr} for player $2$ if and only if it satisfies, for all $ \pi_i \in \Pi_i$,
  \[
    u^2_i(h, \pi_i^{\ast,2},\pi^{\ast,2}_{j},x ) \ge u_i^2(h,  \pi_i,\pi^{\ast,2}_{j},x),
  \]
  where $(i,j)\in \{(1,2), (2,1)\}$.
Note that in the case that P2's hypothesis $x$ is a distribution over $\Phi$, then the utility is calculated based on the expectation, that is, $u_i^2(h, \pi_i,\pi_j,x) =\sum_{\varphi\in \Phi} x(\varphi) u_i^2(h, \pi_i,\pi_j,\varphi)$.

  The strategy $\pi_1^{\ast,1}$ is \ac{sr} for P1 if and only if it satisfies, for all $ \pi_1 \in \Pi_1$,
  \[
    u_1^1(h, \pi_1^{\ast,1}, \pi_2^{\ast,2}, \varphi_1 ) \ge
    u_1^1(h, \pi_1, \pi_2^{\ast,2}, \varphi_1 ),
  \]
  where $\pi_2^{\ast,2}$ is \ac{sr} for player 2.
\end{definition}

In words, a strategy is called subjectively rationalizable for player $i$ if it is the best response in player $i$'s perceptual game to a strategy of player $j$, which, for player $j$, is the best response to player $i$ in player $j$'s subjective view modeled by player $i$'s perceptual game. 
A pair of \ac{sr} strategies $\langle \pi^{\ast,1}_1, \pi^{\ast,2}_2 \rangle$ is called the best-response equilibrium of the hypergame $\hgame^2(x)$.

In level-2 hypergame, P2's strategy is \ac{sr} if it is rationalizable in P2's perceptual game $\game(x)$. P1's strategy is \ac{sr} if it is the best response to P2's \ac{sr} strategy. 
However, P1's SR strategy may not be consistent with P2's predicted SR strategy for P1. %\todo[inline]{This reads wired, I did not get what this is trying to say. This does not connect to the next sentence. }
This inconsistency can be recognized by P2. When P2 notices the mismatch in the perceptual games, her perceptual game may evolve given new information.

\subsection{Dynamic hypergames on graphs}

% The above hypergame model does not account for evolving misperception in the encounter.
To characterize P2's evolving perceptual game, we introduce an inference function.

\begin{definition}[Inference]~\label{def:inference}Assuming P2 has complete observation on the game plays, a \emph{perfect recall inference} function $\eta\colon X\times \prefplays \to  X$ maps a hypothesis $x \in X$ and an observation (a history) $h \in \prefplays$ to a new hypothesis $x' =\eta(x,h)\in X$.
\end{definition}

We introduce a \emph{transition system of P1's level-1 hypergame} to capture simultaneously the changes in game states given players' actions and the evolving perceptual game of P2.

\begin{definition}[Transition System of P1's Level-1 Hypergame]
  \label{def:hypergame_graph}
  Given the transition system $TS = \langle S, A, P, s_0, \calAP, L \rangle$, the \ac{dfa} $\calA = \langle Q, \Sigma, \delta, \iota, F \rangle$ that corresponds to P1's \ac{scltl} specification, and P2's hypotheses space $X$, the transition system of P1's level-1 hypergame is a tuple
  \[
    {\cal H} = \langle V, A, \Delta, (s_0, h_0, q_0, x_0),  {\cal F}  \rangle,
  \]
  where the components of hypergame transition system are defined as follows,
  \begin{IEEEitemize}
    \item $V =S\times \prefplays \times Q \times {X}$ is the set of states. Every state $v =(s, h, q, x) \in V$ has four components:
          \begin{IEEEitemize}
            \item $s$ is the state.
            \item $h \in \prefplays $ is a history  terminating in state $s \in S$.
            \item $q \in Q$ is the automaton state for keeping track of P1's progress towards satisfying $\varphi_1$.

            \item $x \in X$ represents the hypothesis of P2 given the history $h$.
          \end{IEEEitemize}

    \item $A$ is the set of joint  actions.

    \item $\Delta\colon V \times A \to \dist V$ is a probabilistic transition function defined as follows. Consider $v= (s,h, q, x)$ and $v' = (s', h a s',q',x')$, where $h a s'$ is the history $h$ appended with the new action $a$ and state $s'$,
          \begin{align*}
            \Delta(v' \mid v, a) = & P(s' \mid s,a)\indicator(\delta(q, L(s')) = q') \\
                                &\indicator(\eta(x, h a s') = x'),
            \end{align*}
          where $\indicator(E)$ is the indicator function that returns $1$ if the statement $E$ is true, and $0$ otherwise.
    \item $(s_0,h_0, q_0, x_0)$ is the initial state that includes the initial state in the transition system $TS$, the current history that consists of the initial state only, \ie, $h_0 = s_0$, $q_0=\delta(\iota, L(s_0))$, and the initial hypothesis $x_0$.

    \item ${\cal F} = S\times \prefplays\times F \times {X}$ is the set of final states for P1.%, where $F$ is the set of accepting states of the automaton $\cal A$.

  \end{IEEEitemize}
\end{definition}

The transition function is understood as follows: Given a history $h$ ending in the current state $s$ and a joint action $a\in A$, the probability of reaching the next state $s' $ is determined by $P(s' \mid s, a)$ in the hypergame transition system. Upon reaching $s'$, P2 updates her hypothesis to $x' = \eta(x, h a s')$ (here we assume the entire history is used for this update). Also, the transition in the specification automaton is triggered to reach state $q' = \delta(q, L(s'))$ given the labeling of the new state $s'$. % Unless otherwise noted, we refer to the transition system of P1's level-1 hypergame as a dynamic hypergame.

Given P2's perceptual game evolving given the history and the inference function, P2 employs a \emph{behaviorally subjectively rationalizable} strategy. A strategy is behavioral if it depends on the perceptual game when the action is taken \cite{huang2020dynamic}. We define the behaviorally subjectively rationalizable strategy as follows.

\begin{definition}[Behaviorally Subjectively Rationalizable (BSR) Strategy]
  A strategy $\pi_2^{B,2}\colon \prefplays \to \dist {A_2}$ is \ac{bsr} for P2 if
  \[
    \pi_2^{B,2}(h) = \pi_2^{\ast,2} (x, h),
  \]
  where $x=\eta(x_0, h)$,
  and $\pi_2^{\ast,2}(x,\cdot) \colon \prefplays \to \dist {A_2}$ is a subjectively rationalizable strategy for P2 in the hypergame $\hgame^2(x)$.
\end{definition}

\begin{remark}
  When $X$ is finite, the hypergame transition system has a countably infinite set of states. This is because a history can be of a finite but unbounded length. The entire history is maintained as part of the state due to the general definition of the inference mechanism.
\end{remark}

\subsection{Synthesizing P1's Deceptive Strategy}

Given that P2 uses a \ac{bsr} strategy, P1 can play deceptively by influencing P2's hypothesis so that P2's actions given her hypothesis can be advantageous for P1. We leverage the hierarchy of reasoning in level-2 hypergames and develop a two-step approach: In the first step, for each state $v= (s, h, q, x)$, we solve P2's \ac{bsr} strategy using the static hypergame $\hgame^2(x)$. In the second step, we incorporate P2's \ac{bsr} strategies into the transition system in Def.~\ref{def:hypergame_graph}. 

This second step is nontrivial as in some games, there are multiple \ac{sr} strategies for P2 in hypergame $\hgame^2(x)$. We consider a special case when only one equilibrium exists in the P2's perceptual game $\game(x)$, for each $x \in X$. % This may also correspond to situations when there are multiple equilibria but the player follows one equilibrium according to some convention.

We present the solution of P1's strategy for a class of dynamic hypergames that satisfy the following assumptions:
\begin{assumption}
  \label{assume:subclass}
  This subclass of hypergames satisfies the following assumptions:
  \begin{IEEEenumerate}
    \item The hypothesis space $X$ is discrete and finite.
    \item The inference function $\eta$ has a finite domain. That is, the set of histories are grouped into a finite set  of \emph{equivalence classes} (see the formal definition next).
    \item For any $x\in X$, the strategy of P2 in a Nash equilibrium for a given static hypergame $\game(x)$ is unique.
      \item For any $x\in X$, the strategy of P2 in a Nash equilibrium for a given static hypergame $\game(x)$ is   memoryless.
%    \item The strategy of P2 in a best-response equilibrium for a given static hypergame $\hgame^2(x)$ is memoryless, that is $\pi_2^{\ast,2}(x) \colon S \to \dist {A_2}$.
  \end{IEEEenumerate}
\end{assumption}

\begin{definition}[Inference-equivalent histories]
Given an inference function $\eta\colon X \times \prefplays \to X $ and a hypothesis $x$, two histories $h_1$ and $h_2$ are to be said $(\eta,x)$-equivalent if $\eta(x,h_1)= \eta(x,h_2)$ and for any $h'\in (A\times S)^+$, $\eta(x,h_1 h') = \eta(x, h_2 h')$. The set of histories equivalent to $h\in \prefplays$ given hypothesis $x$ is denoted by $[\![h]\!]_{x}$. If the equivalence between histories can be defined to be independent of the current hypothesis, that is,  \emph{for any} pair of hypotheses $x, x'\in X$, if $h_1,h_2$ are $(\eta,x)$-equivalent, then $h_1,h_2$ are also $(\eta, x')$-equivalent, then we say that the  two histories $h_1$ and $h_2$ are \emph{$\eta$-equivalent}. The set of  histories $\eta$-equivalent to $h\in \prefplays$ is denoted by $[\![h]\!]$.
\end{definition}
% It is noted that if two histories are $\eta$-equivalent, then it must be  $(\eta,x)$-equivalent for a given inference $\eta$ and for any hypothesis $x\in X$.  Thus, $\eta$-equivalence relation is a special case of $(\eta,x)$-equivalence relation.

When the set of  $\eta$-equivalent (or $(\eta,x)$-equivalent) histories is finite and the set $X$ is finite,  the inference $\eta$ has a finite domain and can be equivalently expressed using a Moore machine \cite{moore1956gedanken} with inputs  histories and outputs hypotheses. The set of states in the Moore machine is the set of equivalent classes.
We omit the details in constructing the Moore machine and directly incorporate the input-output relations to construct the decision-make problem for solving P1's deceptive strategy.

\begin{definition}
  \label{def:game2mdp}
  Under Assumption~\ref{assume:subclass}, the deceptive strategy for P1 in the dynamic hypergame
  ${\cal H} = \langle V, A, \Delta, (s_0, h_0, q_0, x_0),  {\cal F}  \rangle$ can be obtained by solving the following \ac{mdp} with a reachability objective,
  \[
    \tilde {\cal H} = \langle \tilde V, A_1, \tilde \Delta, (s_0, [\![h_0]\!]_{x_0}, q_0, x_0), \calF \rangle,
  \]
  where
  \begin{IEEEitemize}
    \item $\tilde V $ is a finite and discrete set of states. Each state $\tilde v = (s, [\![h]\!]_x, q, x)$ consists of a state $s$, the partition that includes the history classes given the $(\eta,x)$-equivalence relation, a state $q$ in the \ac{dfa}, and a hypothesis $x$ of P2.
    \item $\tilde \Delta \colon \tilde V \times A_1 \to \dist {\tilde V}$ is defined as follows:
          For any state $\tilde v = (s, [\![h]\!]_x, q, x) $, if $q \equiv q_\sink$---the absorbing state in the \ac{dfa} $\calA$, then state $\tilde v$ is a sink/absorbing state. 

          Given $\tilde v_1 = ( s_1, [\![h_1]\!]_{x_1}, q_1,x_1)$ with $q_1\ne q_\sink$, $a_1\in A_1$, and $\tilde v_2  = (s_2,  [\![h_2]\!]_{x_2}, q_2,x_2)$ and  $h_1 (a_1,a_2) s_2 \in [\![h_2]\!]_{x_2}$, then
          \begin{align*}
            \tilde \Delta(\tilde v_2\mid \tilde v_1 , a_1) = &
            \sum_{a_2 \in A_2}\pi_2^{\ast,2}(a_2\mid  x_1, s_1) \\ & P(s_2 \mid s_1, (a_1,a_2)) \indicator(\delta(q_1, L(s_2)) = q_2).
            \end{align*}
    where $\pi_2^{\ast,2}(a_2\mid x_1, s_1)$ is the probability of P2 selecting action $a_2$ given her current hypothesis $x_1$ and the current state $s_1$. That is, P2 uses a \ac{bsr} strategy.
    \item $(s_0, [\![h_0]\!]_{x_0}, q_0, x_0)\in \tilde{V}$ is the initial state, given $(s_0,h_0, q_0, x_0)$  is the initial state in the transition system ${\cal H}$.
    \item $\tilde \calF = \{ (s,  [\![h]\!]_x, q, x) \in \tilde V\mid q\in F\}$ is the set of final states for P1, where $F$ is the set of final states of \ac{dfa} $\calA$. The states in $\calF$ are absorbing.
  \end{IEEEitemize}
\end{definition}

The size of state space in the \ac{mdp} is $O(\abs{S}\times N \times \abs{Q}\times \abs{X})$, where $N$ is the number of $(\eta, x)$-equivalent classes of histories in the game. The size of the action space in the \ac{mdp} is $O(A_1)$.
By construction, if a path $\rho$ in the \ac{mdp} visits $\tilde \calF$, then P1 satisfies the \ac{ltl} formula $\varphi_1$. Thus, to maximize the probability of satisfying P1's specification, P1 is to maximize the probability of reaching the set $\tilde \calF$. We can formulate a \ac{ssp} problem assuming each path in the \ac{mdp} by following a policy will terminate in a sink state. Solving the deceptive strategy for P1 is of time complexity polynomial in the size of state space and action space of the \ac{mdp}.

An \ac{mdp} with a reachability objective can be solved using probabilistic model checking algorithms (\cite[Chapter~10.1.1]{Baier2008},\cite{KNP07a}) and existing PRISM toolbox \cite{KNP11}. Given the problem can be of large scale, other approximate dynamic programming solutions of \ac{mdp} can be used to reduce the dimension of decision variables \cite{liTopologicalApproximateDynamic2019}.

\begin{lemma}
  Under Assumption~\ref{assume:subclass} and assuming P1's knowledge about $\eta$ is correct, then
  the optimal policy $\pi_1^\ast \colon \tilde V \to \dist {A_1}$ in the \ac{mdp} $\tilde  {\cal H} $ is P1's subjectively rationalizable strategy in the dynamic hypergame given P2's evolving knowledge.
\end{lemma}
\begin{proof}
 The construction of $\tilde {\cal H}$ assumes that P2 follows the behaviorally subjectively rationalizable strategy. Thus, the optimal strategy in $\tilde {\cal H}$ controls both P1's actions and P2's evolving perceptual game so that P2's actions (predicted by P1) are optimal to P1's objective. Deviating from the strategy cannot gain P1 a better outcome. 
\end{proof}

\begin{remark}
  Assumption~\ref{assume:subclass}-4) is not necessary. If P2's \ac{bsr} strategy is represented using a finite-state controller (also known as finite-memory policies), then we can augment the states in the hypergame transition system in Def.~\ref{def:hypergame_graph}
  with the states in the finite-state controller.
\end{remark}

\subsection{Detecting the Mismatch for Opponent Modeling}
\label{subsec:detecting_inconsistency}

The correctness of P1's \ac{bsr} strategy hinges on the correctness in the modeling of P2's inference and predicted P2's \ac{bsr} strategy. In this section, we develop a method to detect if there is any inconsistency between the actual behavior of P2 observed during online interaction and P1's model of P2's behavior used in the \ac{mdp} $\tilde {\cal H}$. If the method identifies the inconsistency, then it alerts P1 that the deceptive policy $\pi^\ast_{1}$ may not be effective.

Consider a finite  history $h = s_0 a_0 s_1 a_1 \ldots s_n$, we denote the action pairs for P1 and P2 as $a_i = (a_{1,i},a_{2,i})$ for any $0\leq i \leq n-1$. Under the assumption of complete observations, both players can observe the history $h$. By employing the \ac{dfa} $\calA$ corresponding to P1's \ac{ltl} specification $\varphi_1$ and the inference function $\eta$ in the \ac{mdp} $\tilde{\cal H}$, we can transform the history $h$ to an augmented state-action sequence denoted by $\tilde h = \tilde{v}_0 a_0\tilde{v}_1a_1\ldots\tilde{v}_n$, where $\tilde{v}_i = (s_i, [\![h_i]\!]_{x_i}, q_i, x_i)$ and $\tilde{v}_{i+1} = (s_{i+1}, [\![h_{i+1}]\!]_{x_{i+1}}, q_{i+1}, x_{i+1})$, $q_{i+1} = \delta(q_i, L(s_{i+1}))$, $h_{i+1} = h_i a_i s_{i+1}$, $x_{i+1} = \eta(x_i, h_{i+1})$. It is worth noting that for a unique history $h$, there exists a unique augmented state-action sequence $\tilde h$ as the transitions in the \ac{dfa} and inference functions are deterministic, and the entire history up to step $i+1$ is used to compute the $(\eta, x_{i+1})-$equivalent histories at step $i+1$. The detection problem reduces to: Given the transition system $TS$, a predicted inference function $\eta$ for P2 and P2's \ac{bsr} strategies for a set of hypotheses $X$, how likely is the observation $\tilde h$  generated by our predicted model of opponent?
If the answer is yes, then there is no mismatch in the model. 

We employ  likelihood ratio test \cite{severini2000likelihood} to answer this question. We have two hypotheses---$H_0:$ the data is generated by our predicted model of P2; and $H_1:$ the data is not generated by our prediction. The goal is to test which hypothesis is a good fit for the data. %We consider a simpler case and assume our predicted model of inference $\eta$ is correct for P2, and want to test if P2  deviates from the predicted \ac{bsr} strategies.
Given P1's action sequences, the state sequences, and the \ac{mdp}  $\tilde{\cal H}$ from Def.~\ref{def:game2mdp}, the predicted P2's  policy, the likelihood of P2's action sequences is computed by
\[
  L_0= L(a_{2, 0}a_{2, 1}\ldots a_{2, n-1} \mid \pi_2^{\ast,2}) = \prod_{i=0}^{n-1}\pi_2^{\ast,2}(a_{2, i}\mid x_i, s_i),
\]
where $s_i=\tilde v_i[1]$ and $x_i  = \tilde v_i[4]$ are the first and last components of $\tilde v_i$.

At the same time, we obtain an estimate of P2's strategy from history as
\[
  \hat \pi_2 (a_2\mid \tilde v) =\left\{ \begin{array}{ll}
    \frac{ \#\indicator( \tilde v, a_2)}{\#
    \indicator(\tilde v)} & \text{ if } {\#
      \indicator(\tilde v)}\ne 0,\\
    \uparrow              & \text{ undefined otherwise.}
  \end{array}\right.
\]
where $\#\indicator(\tilde v,a_2)$ is the number of
times P2 selects action $a_2$ given the
current state $\tilde v$, and $\#\indicator(\tilde v)$ is the number of times the state $\tilde v$ is visited. For unseen state-action pairs, we do not estimate the policy given the pair as it won't be used in the  test. Based on the maximum likelihood estimate of P2's strategy from history, we have
\[
  L_1 = L(a_{2, 0}a_{2, 1}\ldots a_{2, n-1} \mid \hat \pi_2) = \prod_{i=0}^{n-1}\hat \pi_2(a_{2, i}\mid x_i, s_i).
\]
The likelihood ratio is computed as
\[
  \lambda = L_0/L_1.
\]
We conduct a likelihood ratio test and
calculate $\chi^2_n= -2 \ln \lambda$, which is an approximate Chi-square distribution of $n$ degree of freedom, and $n$ is the number of parameters $\{ \hat \pi_2(a_2\mid \tilde v)\mid\#\indicator(\tilde v,a_2)\ne 0\}$ estimated with maximum likelihood estimation. By selecting a confidence level $\alpha$, we reject the null hypothesis $H_0$ if $\chi^2_n$ is larger than the Chi-square percentile  with $n$ degrees of freedom given the level $\alpha$.

\begin{comment}
 Given that the sequence of P2's hypothesis $x_i$, $i=0,\ldots, n$ is not observable by P1, we cannot evaluate whether our predicted inference function $\eta$ of P2 is the exact inference function used by P2. To detect the mismatch in both P2's strategies and inference mechanism, we would need to estimate P2's policy from the observed history and compare that with the null hypothesis. As P2 can select finite-memory strategy modeled by a finite-state automaton with unknown structure, the policy cannot be learned from positive data (that is, what P2 does). To learn a finite-state automaton without the structure, both positive data and negative data are required~\cite{Neider2019}.

However, learning the true policy of P2 is not our goal. For deception purpose, as long as P2's actual behavior fits our predicted behavior of P2 (null hypothesis $H_0$), we consider the deception is effective.

\begin{remark}
  The detection algorithm monitors P2's behavior during online interaction. However, this likelihood ratio test can easily generalize to multiple histories during repeated interactions. P1 may also use this test to self-validate his model of P2 before carrying out important missions.
\end{remark}
\end{comment}

%% file: sections/case_study.tex
In this section, we present an example to illustrate how this solution concept of dynamic level-2 hypergames gives rise to deceptive strategies for a class of opponent models that satisfy Assumption~\ref{assume:subclass}. This case study includes an inference function for P2 (Sec.~\ref{sec:inference}) based on sliding-window change detection, the construction of \ac{mdp} $\tilde {\cal H}$ for planning the deceptive optimal strategy for P1, with application to robot motion planning with temporal logic objectives in an adversarial environment.

\subsection{Inference with sliding-window change detection}
\label{sec:inference}
We introduce a class of inference algorithms based on change detection in \ac{mc}~\cite{Polansky2007}.
Given P2's finite hypothesis space $X$, P2 can construct a set of games $\{\game(x)\mid x\in X\}$. For each game $\game(x)$, it is assumed that there is an unique equilibrium $\langle \pi_1^\ast(x), \pi_2^\ast(x)\rangle$ where $\pi_i^\ast\colon \prefplays \times X \rightarrow \dist {A_i}$ is a mixed strategy for player $i$. This equilibrium induces, from the concurrent game graph $\game$, a probability measure $\Pr^{\langle \pi_1^\ast(x), \pi_2^\ast(x)\rangle}$ over histories in $\game$. For simplicity in notation, we denote $\Pr^{\langle \pi_1^\ast(x), \pi_2^\ast(x)\rangle}$ by $\Pr^x$.

When P2's current hypothesis is $x$, P2 can detect a change from $x$ to some $x' \in X$ using a sliding-window change detection algorithm, based on the \ac{cusum} statistic~\cite{basseville1993detection}.
First, we are given a data point in forms of history (joint state-action sequence) $h = s_0 a_0 s_1 \ldots s_n$ and a nominal model $x_0$. We denote the interval of a time window of size $m+1$ as $[k, k + m]$, and the history within this time window is $s_k a_k \ldots s_{k+m} a_{k+m} s_{k+m+1}$. Second, we denote the $i$-th observation of the transitions \emph{within the time window} as $y_i = (a_{k+i-1}, s_{k+i})$ for any $i > 0$. When $i=0$, the $0$-th observation within the window is $y_0 = (s_{k})$.
Intuitively, given a data and a nominal model $x_0$, the sliding-window change detection algorithm uses a subsequence of history over a time window and detects if a change has occurred in the model that generates the data during this time window. Specifically, for each hypothesis $x\in X$ and a nominal model $x_0$, the algorithm computes the log-likelihood ratio,
\begin{align}
  R_j^x =\sum_{i=0}^j r_i^x,
\end{align}
where $r_i^x= \ln \frac{\Pr^{x}(y_i)}{\Pr^{x_0}(y_i)}$, and $\Pr^{x}(y_i)$ (resp. $\Pr^{x_0}(y_i)$) is the probability of observing the transition given the probability measure $\Pr^x$ (resp. $\Pr^{x_0}$).

The change detection lies in the difference between the log likelihood ratio and its current minimum value. The \ac{cusum} score is given by
\begin{align}
  Z_k^x  = R_k^x - \min_{1\le j\le k} R_j^x.
\end{align}
Recursively, the \ac{cusum} score is updated for each hypothesis $x \in X$ as
\begin{align}
  Z_k^x = \max\{0, Z_{k-1}^x + \ln \frac{\Pr^{x}(y_k)}{\Pr^{x_0}(y_k)} \},
\end{align}
where $Z_0^x= 0$.

A change is detected at time $t$ when the score of at least one model, say $Z_k^{x}$, exceeds a user-defined constant threshold $c>0$. Formally, the \emph{time of change} is given by
% \[
%   t = \min\{k:  \exists x \in X,\; Z_k^x \ge c\}.
% \]
% \todo[inline]{
\[
  t = \min\{k\mid  \exists x \in X,\; Z_k^x \ge c\}.
\]
% }
Once a change is detected, the algorithm sets the nominal model to be the current predicted model, disregards the history until the change, and keeps running the online change detection given new observation from the change point onwards. In the case when multiple models maintain similar \ac{cusum} scores, we select one model based on some domain-specific heuristics or at uniformly random.

\begin{lemma}
  Given a sliding-window change detection inference $\eta\colon X\times \prefplays \rightarrow X$ with window size $m+1$ and a finite hypothesis space $X$, two histories $h_1, h_2$ are $(\eta,x)$-\emph{equivalent} if they share the same suffix~\footnote{For a word $w = \sigma_1 \sigma_2 \ldots \sigma_n$, a suffix of $w$ is a word $v$ of the form $\sigma_i \sigma_{i+1} \ldots \sigma_n$, where $1 \leq i \leq n+1$.} of length $m+1$.
\end{lemma}
The proof is based on the property of the change detection and thus omitted. % Intuitively, from the time window of size $m-1$, the two history $h_1$ and $h_2$ share the same state-action sequence.

\subsection{Deceptive Planing with Temporal Logic Objective}
\label{subsec:blocking}
\begin{figure}[!htb]
  \begin{subfigure}[b]{.3\linewidth}
    \includegraphics[width=1\linewidth]{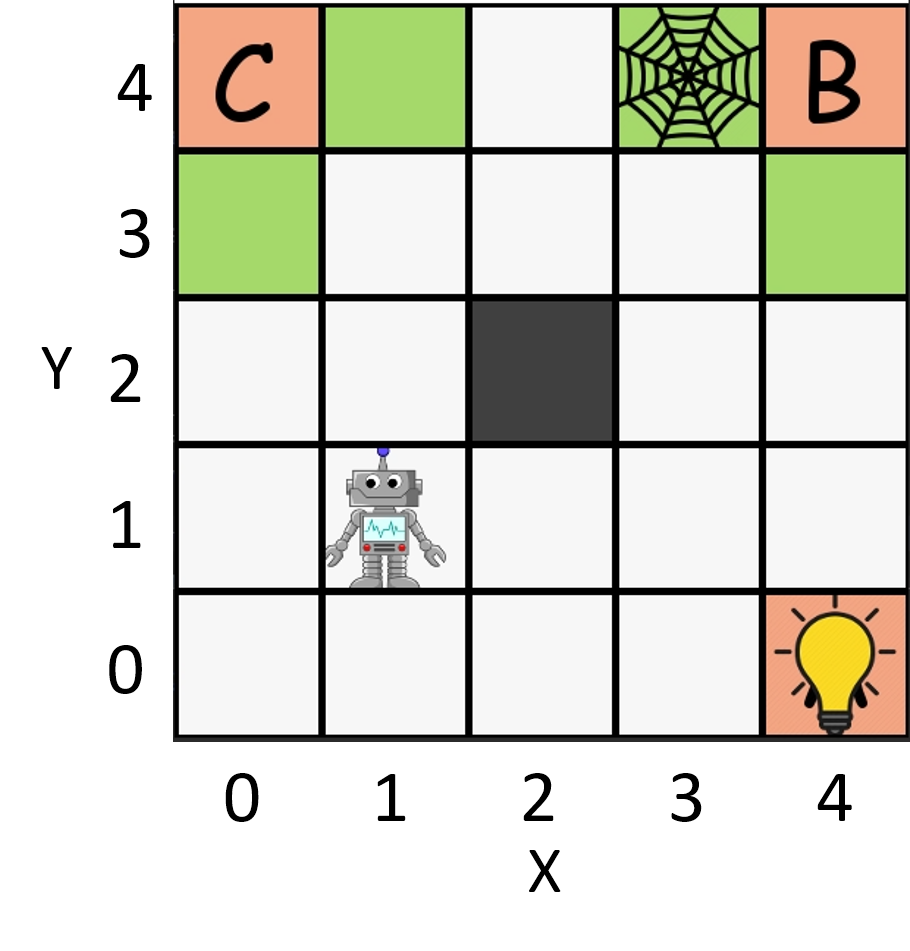}
    \caption{ }
    \label{fig:grid_world_1}
  \end{subfigure}
  \hfill
  \begin{subfigure}[b]{.3\linewidth}
    \includegraphics[width=1\linewidth]{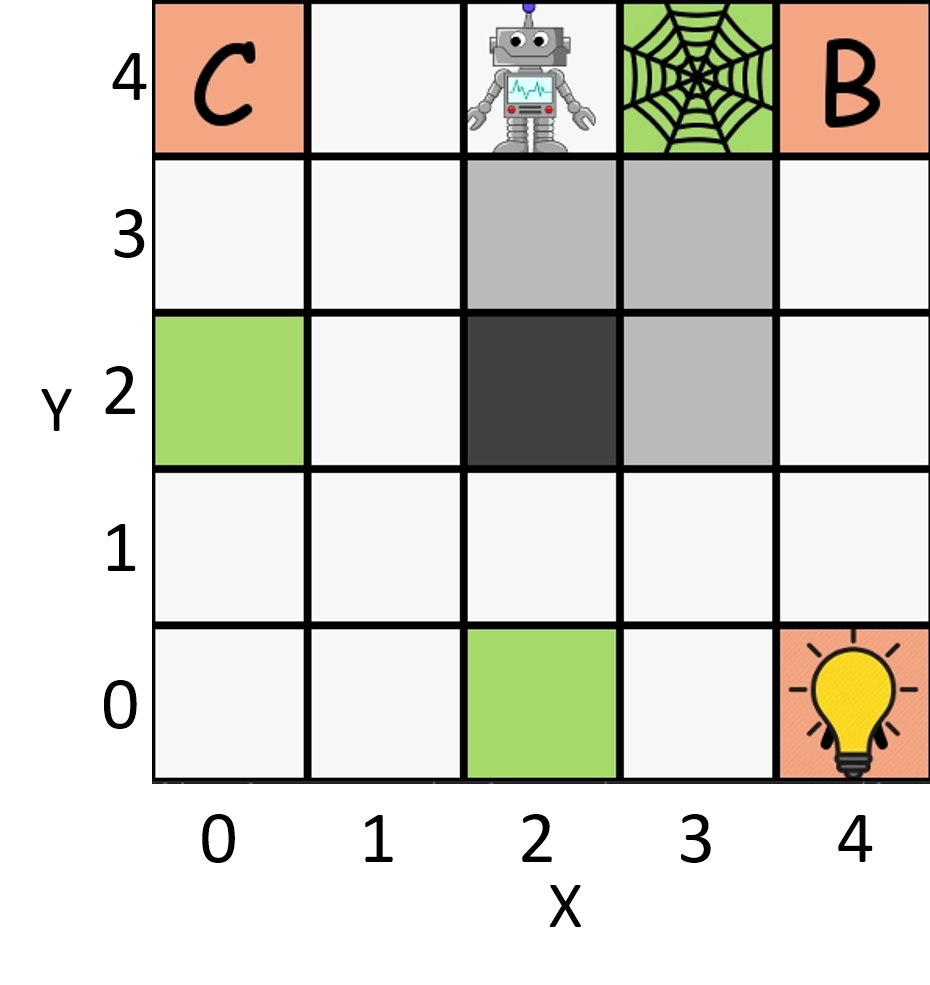}
    \caption{ }
    \label{fig:grid_world_2}
  \end{subfigure}
  \hfill
  \begin{subfigure}[b]{.3\linewidth}
    \includegraphics[width=1\linewidth]{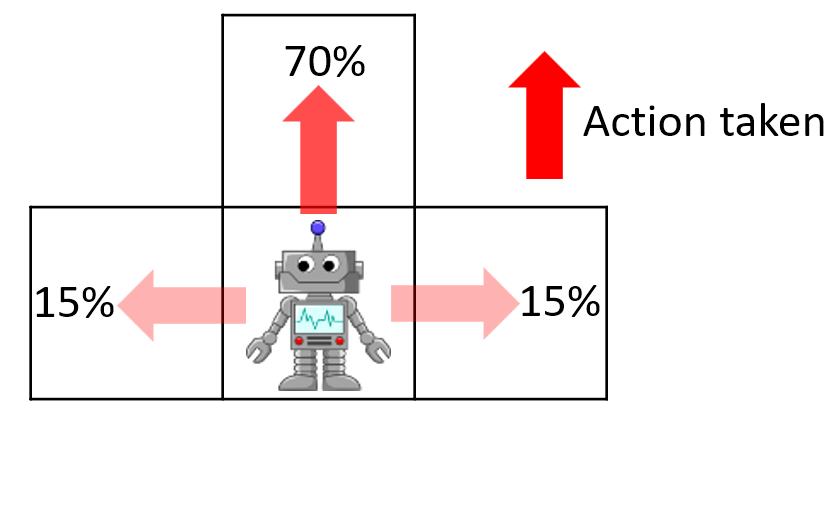}
    \caption{ }
    \label{fig:dynamics}
  \end{subfigure}
  \caption{(a): $world_1$'s initial configuration for P1 and P2. (b): $world_2$'s initial configuration for P1 and P2. Cells colored in grey are walls. Bulb indicates initial P2's prediction. (c): Robot's dynamics when action "up" is taken.}
  \label{fig:two_grid_worlds}
\end{figure}

We consider two examples inspired by security games, referred to $world_1$ (Fig.~\ref{fig:grid_world_1}) and $world_2$ (Fig.~\ref{fig:grid_world_2}).
In both worlds, a robot (P1) is to visit several regions of interests (labelled $A, B, C$ and colored in red) according to a temporal ordering, and an observer (P2) can reallocate traps in cells colored in green. Both games are concurrent: When P1 selects an action to move, the P2 selects an action to set traps. When P1 enters the cell where P2 allocates the trap to that cell, P1 is trapped. The game terminates in two ways: a) P1 is trapped; b) P1 completes its task.

\begin{figure}[!htb]
  \centering
  \begin{tikzpicture}[->,>=stealth',shorten >=1pt,auto,node distance=3cm,scale=.7,semithick, transform shape]
    \tikzstyle{every state}=[fill=black!10!white]
    \node[initial, state]   (1)                                     {$q_0$};
    \node[state]            (2) [below right of =1]                 {$q_1$};
    \node[state]            (3) [below left of =1]                  {$q_2$};
    \node[state]            (4) [below of =2]                       {$q_3$};
    \node[state, accepting] (0) [below of =3]                       {$q_4$};

    \path[->]
    (0) edge[loop left]     node{$\BoolTrue$}      (0)

    (1) edge                node[left,near start]{$\mbox{obs} \vee B$}      (4)
    (1) edge                node[above left]{$A$}               (3)
    (1) edge                node{$C$}               (2)

    (2) edge                node{$\mbox{obs}$}             (4)
    (2) edge                node[near end]{$A$}     (0)
    (2) edge[loop above]    node{$B \vee C$}        (2)

    (3) edge                node[above right,near start]{$\mbox{obs} \vee B$}         (4)
    (3) edge[loop above]    node{$A$}               (3)
    (3) edge                node{$C$}     (0)

    (4) edge[loop right]    node{$\BoolTrue$}       (4)
    ;
  \end{tikzpicture}
  \caption{The task automaton with $5$ states and $12$ edges corresponds to $\varphi_1$, where $Q = \{q_i \mid i =0, 1, 2, 3, 4\}$.}
  \label{fig:example_dfa}
\end{figure}
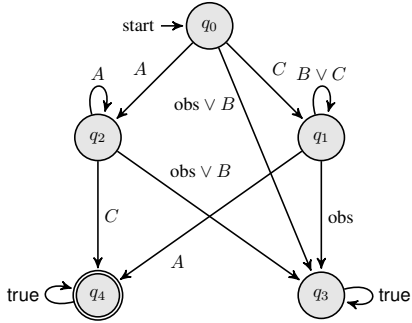

Formally, we describe P1's task by the formula
\[
  \varphi_1=  (\neg \mbox{obs} \until A) \land (\neg (B \vee \mbox{obs}) \until C)).
\]
That is, the robot needs to visit $A$ and $C$ without reaching obstacles. Before visiting $C$, the robot cannot visit $B$. The corresponding \ac{dfa} is drawn in Fig.~\ref{fig:example_dfa}.

P1 can move in four compass directions, and P1's dynamics is plotted in Fig.~\ref{fig:dynamics}. The grid world is surrounded with bouncing wall, \ie, if P1 hits the wall, it bounces back to its previous cell. The black cells in the grid world is a static obstacle, labeled by $\mbox{obs}$.

P2 can reallocate the traps (\ie, dynamic obstacles) to a subset of cells colored in green in $world_1$ and $world_2$.  P2 can only use $\ell$ traps with $n$ possible trap locations. Thus, the number of actions for the P2 is ${n \choose{\ell}}$, \ie, choose $\ell$ out of $n$. Every time P2 resets the location of any trap, it must wait at least $k$ time steps to be able to reallocate any trap again.  In the example of $world_1$, we select $n=4$, $\ell=1$, and let $k=0$; In the example of $world_2$, we select $n=3$, $\ell=1$, and let $k$ to be a variable.

In both examples: $world_1$ and $world_2$, the asymmetrical information is as follows:
\begin{IEEEitemize}
  \item P1 knows the complete task $\varphi_1$.
  \item P2 does not know the complete task $\varphi_1$.
\end{IEEEitemize}
We refer to this situation as \emph{asymmetric information} case. On the other side, if P2 knows P1's complete task, then we refer to that as \emph{symmetric information} case. In the \emph{asymmetric information} case, P2 has a hypothesis space $X = \{\neg \mbox{obs} \until \phi \mid \phi \in \{A, B, C\}\}$.

\subsubsection{Different behaviors under asymmetric and symmetric information in $world_1$}
We compare P1's task completion rate between asymmetric information case and symmetric information case. %We show that P1 can leverage information asymmetry to achieve a higher probability of completing the task.

In asymmetric information case, for each $x \in X$, P2 solves a Stackelberg/leader-follower game and decides a trap configuration against the best response of P1 in game $\game(x)$. Let $A_2$ be the set of different configurations of traps. The strategy of P2 is to maximize the following objective function
\[
  \pi_2^{B,2}(s) = \arg\min_{a_2 \in A_2}  \max_{\pi_1}\Pr^{\pi_1}(h \models x\mid s, a_2), \text{for all } s \in S,
\]
where $\Pr^{\pi_1}(h \models x \mid a_2)$ is the probability of P1 satisfying the formula $x$ given P2's action (trap configuration) $a_2$. For instance, if $x = \neg \mbox{obs}\until B$ and robot is at the $(2,4)$, then P2's optimal action is to allocates the trap to the green cell right to robot, that is $(3,4)$ in Fig.~\ref{fig:grid_world_1}. For each hypothesis $x \in X$ and state $s \in S$, P2 solves the optimal trap allocation action $a_2$ and also computes the best response of P1, $\pi_1$, that achieves the maximum probability of satisfying $x$ from $s$. The joint strategy profile provides the probability measure $\Pr^{x}$ for P2 to  infer the subgoal of P1 using the sliding window change detection.

For the configuration of $world_1$, we evaluate different window sizes and find sliding-window size $m+1 = 2$ achieves a good trade-off in space complexity and accuracy in prediction. In case we choose an unsuitable window size, our method in subsection~\ref{subsec:detecting_inconsistency} enables P1 to detect the incorrect prediction of P2's behavior model caused by unsuitable windows size and increase the window size. %If the consistency is identified, then it means that we have a mismatch between P2's modeled behavior and the actual P2's behavior model.
Assuming P2 can reallocate traps anytime with $\pi_2^{B, 2}$, we can construct an \ac{mdp} $\tilde{\cal H}$ and solve for P1's optimal deceptive strategy denoted by $\pi^{\ast}_1$. % Then we get a \ac{bsr} strategy profile $\langle \pi_1^{\ast}(\tilde{{\cal H}}), \pi^{B,2}_2(x) \rangle$.

In symmetric information case, P1 and P2 both have exact knowledge of task specification $\varphi_1$, and P1 wants to maximize the P1's probability of finishing the task; P2 wants to minimize the P1's probability of finishing the task. We denote the Nash Equilibrium strategy profile by $\langle \pi_1^\NE, \pi_2^\NE \rangle$. Where the Nash Equilibrium strategy profile is to solve the following objective function
\[
  \langle \pi_1^{\NE}, \pi_2^\NE \rangle = \arg\min_{\pi_2 \in \Pi_2}  \max_{\pi_1 \in \Pi_1}\Pr^{\pi_1, \pi_2}(h \models \varphi_1).
\]
%Given the transition system $TS$ and the Nash equilibrium strategy profile $\langle \pi^\ast_1, \pi_2^\ast \rangle$, we can obtain an induced interaction between P1 and P2, where P1 follows policy $\pi^\ast_1$ and P2 follows policy $\pi_2^\ast$.

\begin{table}[!htb]
    \caption{For $world_1$, completion rates for P1 in asymmetric information case and symmetric information case.}
    \label{tab:winning_rate}
    \centering
    \begin{tabular}{lllr}
        \hline
   Info&   P1 Policy                      & P2 Policy         & Completion rate (P1) \\ \hline
Asymmetric &      $\pi_1^\ast $ & $\pi^{B, 2}_2$ & \textbf{97.23\%}                         \\
     Symmetric & $\pi^\NE_1$                   & $\pi^\NE_2$      & 33.55\%\\
      \hline
    \end{tabular}%
\end{table}

In Table~\ref{tab:winning_rate}, we list two P1's completion rates of its task specification: one for asymmetric information case and one for symmetric information case case. From the Table~\ref{tab:winning_rate}, it indicates that under asymmetrical information, by following the deceptive strategy given P2 plays \ac{bsr} strategy, P1 has much higher probability of satisfying the specification than that of the case by following the Nash equilibrium strategy profile. % This difference on the completion rates is referred as \emph{value of information}.
The deceptive strategy leverages this information asymmetry to lead P1 to achieve higher probability of finishing it's specification. We provide a short video \footnote{\url{https://www.dropbox.com/s/7xxrq6mzm9umvva/07162020.mpeg?dl=0}} to demonstrate the difference on P2's behaviors in asymmetric information case and symmetric information case.

Next, we want to understand whether a delay in reallocating  traps for    P2 would effect the completion rate of P1. However, in the $world_1$ example, %delay only matters when P1 approaches the $B$ or $C$. Same argument applies to the detection of model mismatch, the mechanism may identify the inconsistency at the last step. Moreover, 
%from $world_1$ example, 
we observed in experiments that any delay on reallocation could easily lead P1 to complete its task. Based on these observations, we construct another example $world_2$, and evaluate the complete rates for every $k$ steps of delay and correctness of model mismatch in this example $world_2$.

\subsubsection{Reallocation every $k$ steps of delay in $world_2$}
\begin{figure}[!htb]
  \centering
  \includegraphics[width=\linewidth]{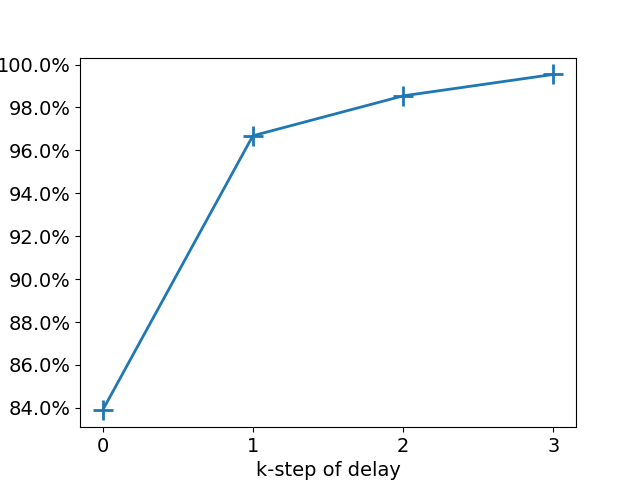}
  \caption{The task completion rates of P1 given P2 with $k$-step delay in reallocating traps, for $k = 0, 1, 2, 3$.}
  \label{fig:values_initial}
\end{figure}

In this example, we assume that P2 is restricted to only reallocate the trap after $k$ steps since the last reallocation, where $k $ is an integer. P1 is aware of P2's delay $k$ and synthesize the deceptive strategy.
 Figure~\ref{fig:values_initial} shows the completion rate of task (values of in initial state $(2, 4)$ in Fig.~\ref{fig:grid_world_2}) under different steps of delay up to $k = 3$. The results indicates that with the increase of steps of delay, the probability of completing the task increases and P1 exploits P2's delay and lack of information. 

\subsubsection{Detection of Model mismatch in $world_2$} We use experiments in the configuration $world_2$ to demonstrate the correctness of the detect mechanism, that is, to identify whether there is a deviation from the predicted opponent modeling of P2.  We set the significance level $\alpha = 0.05$. If the likelihood of observed action sequences is smaller than or equal to $0.05$, we reject the null hypothesis: the data is generated by our predicted model of P2.

\begin{figure}[!htb]
  \centering
  \includegraphics[width=\linewidth]{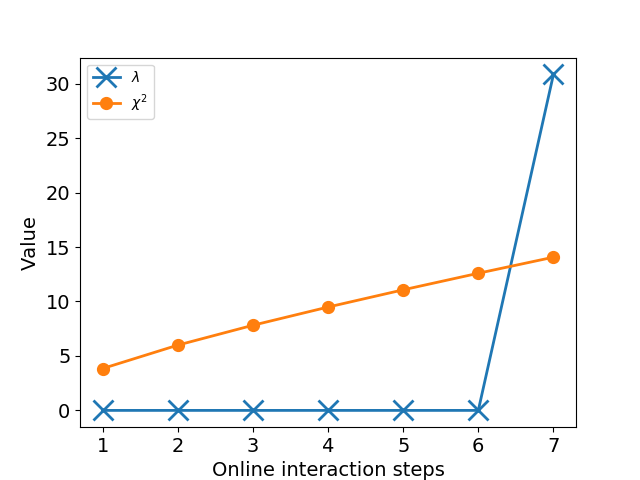}
  \caption{The likelihood ratio $\lambda$ for online interaction between P1 and P2.}
  \label{fig:ratio}
\end{figure}

We consider a case that P1 follows policy $\pi_1^\ast $, and P2 plays her policy predicted  by P1 for first four steps. After first four steps, we let P2 plays a random policy $\pi_2^R$, \ie, $\pi_2^R(a \mid s) = \frac{1}{\abs{A(s)}}$, for all $a \in A(s)$. The mismatch is detected at the $7$-th step of the online interaction, and P1 is alerted that P2 deviates from the predicted policy. We compute $\lambda$ after each step and plot it in Fig.~\ref{fig:ratio}, where we also plot the $\chi^2$. (The reason predicted $\lambda = 0$ is because the predicted policy $\pi_2^{B,2}$ is deterministic.) From  Fig.~\ref{fig:ratio}, we see that at the $7$-th step of online interaction, we have $\lambda > \chi^2$, so we reject the null hypothesis. The degree of freedom in the Chi-square detector is the number of the state-action pairs.

%% file: sections/conclusion.tex
In this work, we propose a solution concept for a class of hypergames to solve deceptive strategies with temporal logic objectives. Our hypergame framework identifies two key components for deceptive planning: Opponent modeling and proactive planning. The general framework can be extended to other class of games with incomplete information. 
It is also important to note that the proposed approach does not generalize easily to partially observable games with two-sided partial observations. This is because that two players may have different observations over the same history and incomplete information about what observation the other player has. The partial observation may potentially make player 1's subjective model of player 2's perceptual game diverges from the actual perceptual game of player 2. 

Building on this work, our future work will focus on adaptive and robust deception for defense. 
In the class of hypergames considered, we have made some assumptions about the inference mechanism and strategies employed by P2. To generalize from deterministic inference to probabilistic inference, the Markov decision process reduced from the game can be continuous, because the hypothesis space $X$ is probabilistic distributions. In addition, robust  Markov decision processes can be incorporated to deal with  mismatches in the opponent model, provided the range of mismatches is known a prior. Robust and adaptive deception is also needed to deal with the case when P2 may have multiple \ac{bsr} strategy, as P1 needs to learn which strategy P2 uses and adapts  its deceptive strategy accordingly. We will also consider practical applications of the deceptive planning for security applications in cyber-physical systems.